\documentclass[]{article}
\usepackage{mathtools,tikz,pgfplots,xcolor,amsmath,comment,graphicx,subcaption}
\usetikzlibrary{math}
\pgfplotsset{compat=1.12}
\parskip=\medskipamount

\definecolor{myred}{RGB}{214,39,40}
\definecolor{mygreen}{RGB}{44,160,44}
\definecolor{myblue}{RGB}{31,119,180}
\definecolor{myorange}{RGB}{254,127,14}

\newcommand{\e}{\mathrm{e}}

\title{A Dynamical Systems Approach to The Fourth Painlev\'e Equation}
\author{Jeremy Schiff \\ Department of Mathematics, \\ Bar-Ilan University, Ramat Gan, 5290002, Israel \\ {\tt schiff@math.biu.ac.il} \and Michael Twiton \\ Department of Mathematics, \\ Bar-Ilan University, Ramat Gan, 5290002, Israel \\ {\tt mtwito101@gmail.com}}

\begin{document}
	
	\maketitle
	\begin{abstract}
	  We use methods from dynamical systems to study the fourth Painlev\'e equation $P_\mathrm{IV}$. Our starting point is the
          symmetric form of $P_\mathrm{IV}$, to which the Poincar\'e compactification is applied. The motion on the sphere at infinity
          can be completely characterized. There are fourteen fixed points, which are classified into three different types. Generic
          orbits of the full system are curves from one of four asymptotically unstable points to one of four asymptotically stable points,
          with the set of allowed transitions depending on the values of the parameters. This allows us to give a qualitative
          description of a generic real solution of $P_\mathrm{IV}$. 
	\end{abstract}
	
	\section{Introduction}
	
	The six Painlev\'e equations are second order differential equations with up to four parameters, that were discovered
        over a century ago, and have been extensively studied since, particularly in the last forty years
        (see, for example, \cite{clarkson2003painleve,bk1,bk2} for references). 
	However, it remains the case that most of the substantial body of
	knowledge about solutions of these equations concerns special
	solutions for special values of the parameters, and there is a
	dearth of knowledge about generic solutions for generic parameter
	values. The aim of this paper is to improve this situation
	for the fourth Painlev\'e equation ($P_\mathrm{IV}$), at least
	for the case of real-valued solutions of a real variable.
	$P_\mathrm{IV}$ is the equation 
	\begin{equation}
		\label{eq:p4}
		\frac{{\mathrm{d}}^{2}w}{{\mathrm{d}z}^{2}}=\frac{1}{2w}\left(\frac{\mathrm{d}
			w}{\mathrm{d}z}\right)^{2}+\frac{3}{2}w^{3}+4zw^{2}+2(z^{2}-\alpha)w+\frac{
			\beta}{w},
	\end{equation}
	with two parameters, $\alpha$ and $\beta$, and we will restrict to the case $\beta \le 0$.
        See \cite{clarkson2008fourth} for an extensive survey of works on $P_\mathrm{IV}$.         
        We will work with the ``symmetric form'' of $P_\mathrm{IV}$, the three-dimensional autonomous dynamical system
	\begin{subequations}
		\label{eq:fsys}
		\begin{align}
			\frac{\mathrm{d} f_1}{\mathrm{d} x}&=f_1 (f_2-f_3)+\alpha_1, \\
			\frac{\mathrm{d} f_2}{\mathrm{d} x}&=f_2 (f_3-f_1)+\alpha_2, \\
			\frac{\mathrm{d} f_3}{\mathrm{d} x}&=f_3 (f_1-f_2)+\alpha_3, 
		\end{align}
	\end{subequations}
	subject to 
	\begin{equation}
		\label{eq:fsum}
		f_1+f_2+f_3=x,
	\end{equation}
	and
	\begin{equation}
		\label{eq:alphasum}
		\alpha_1 + \alpha_2 + \alpha_3 = 1.
	\end{equation}
        The symmetric form of $P_\mathrm{IV}$ was apparently known to Bureau (see \cite{bureau} pp. 115--116),
        but was rediscovered, amongst others, by Adler \cite{adler} 
        and Noumi and Yamada \cite{noumi1998affine,noumi1999symmetries}. 
        If $f_1,f_2,f_3$ is a solution of \eqref{eq:fsys}--\eqref{eq:alphasum} and we set $w(z) = -\sqrt{2} f_1(x)$,
	where $z = x/\sqrt{2}$, then $w(z)$ is a solution of \eqref{eq:p4} with parameter values
	$\alpha = \alpha_3 - \alpha_2$ and $\beta = -2 \alpha_1^2$. Note another symmetric form of $P_\mathrm{IV}$
        was given in \cite{schiff1994backlund}, equation (13)$'$ with $N=3$. 
	
	The symmetric form is particularly useful for discussion of the symmetries and B\"acklund transformations
        of $P_\mathrm{IV}$ \cite{noumi1998affine,noumi1999symmetries,sen2005,sen2006lax}. However, at first glance, its utility
        as a tool to study solutions seems limited.
	From the constraint (3), the dynamical system (2) clearly can have no
	fixed points or periodic orbits, and all orbits must be unbounded. The
	first thing we do in this paper is to apply Poincar\'e compactification
	to the system \eqref{eq:fsys}--\eqref{eq:alphasum}. Poincar\'e compactification is a standard tool in the study
        of polynomial dynamical systems, most heavily used in the
	classification of two-dimensional dynamical systems, see for example \cite{cima1990bounded}.
	Poincar\'e compactification replaces a dynamical system on $\mathbf{R}^n$ by a dynamical
	system on the open unit ball in $\mathbf{R}^n$. After a change of parametrization
	along the orbits, this can be extended smoothly to the closed unit ball,
	with the unit sphere becoming an invariant manifold. Unbounded orbits in
	the original system on $\mathbf{R}^n$ (possibly with finite ``escape times'', i.e.
	diverging in finite time) become bounded orbits in the new system, that
	tend to the ``sphere at infinity'' in infinite time (after the
	reparametrization). 
	
	In the case of the symmetric $P_\mathrm{IV}$ system, we show that  the Poincar\'e compactification has fourteen fixed points on the sphere at infinity. Four of these are attractors (in the sense that nearby orbits inside the sphere converge to them as $t$ tends to $+\infty$), and four are repellers (in the sense that nearby orbits inside the sphere converge to them as $t$ tends to $-\infty$). The remaining six are of mixed type. This holds for arbitrary values of the parameters $\alpha_1,\alpha_2,\alpha_3$. We deduce that a {\em generic} orbit of the Poincar\'e compactification ``starts'' at one of the repellers and ``ends'' at one of the attractors. The only remaining question, with regard to generic orbits, is whether there are orbits between each repeller-attractor pair. Assuming that $\alpha_1,\alpha_2,\alpha_3$ are all nonzero, we show that certain transitions are forbidden, with the rules depending on the signs of $\alpha_1,\alpha_2,\alpha_3$. We illustrate these rules in numerical experiments. 
	
	The previous paragraph concerns the Poincar\'e compactification. To revert to the symmetric form of $P_\mathrm{IV}$ (or to $P_\mathrm{IV}$ itself) we need to take into account the fact that orbits going to (coming from) three of the attractors (repellers) in infinite time $t$ (after the reparametrization), correspond to solutions of symmetric $P_\mathrm{IV}$ going to (coming from) a pole-type singularity in {\em finite} time $x$ (prior to the  reparametrization). However, since these singularities are pole-type, the solutions can be continued past them, corresponding to a concatenation of orbits of the compactification.  Solutions going to (coming from) the fourth attractor (repeller) of the compactification corresponds to solutions of symmetric $P_\mathrm{IV}$ that diverge, but remain finite, as $x\rightarrow +\infty$ ($-\infty$). Thus we obtain a picture of the generic solution of symmetric $P_\mathrm{IV}$ on the real axis in the case where $\alpha_1,\alpha_2,\alpha_3$ are all nonzero. There is a certain way in which the solution can diverge as $x\rightarrow-\infty$ and a certain way in which the solution can diverge as $x\rightarrow+\infty$. Otherwise, the solution consists of transitions between one kind of singular behavior to another. These are subject to rules on which kinds of transitions are allowed, depending on the signs of $\alpha_1,\alpha_2,\alpha_3$. (We emphasize this is the description of generic solutions; there are also non-generic solutions with exceptional behaviors.) 
	
	The structure of this paper is as follows. In Section \ref{sec:compactificationAndEPA} we present the Poincar\'e compactification of symmetric $P_\mathrm{IV}$, its fixed points, and the linearizations of the flow at the fixed points. Only six of the fixed points are hyperbolic, and for the non-hyperbolic points stability cannot be immediately determined from local linearized flows. Before delving more deeply into this, in Section \ref{sec:infsphere} we integrate the flow on the sphere at infinity. Remarkably, this restricted flow exhibits a conserved quantity which we find explicitly. The conserved quantity, however, is singular on a great circle, allowing the six hyperbolic fixed points that lie on this great circle to be nodes (which are prohibited in a two-dimensional system with a regular conserved quantity). In Section \ref{sec:4.1} we continue the study of stability of the fixed points, and reach the central conclusion already stated above: That a generic orbit of the Poincar\'e compactification starts at one of the four repellers on the sphere at infinity  and ends at one of the four attractors on the sphere at infinity. In Section \ref{sec:4.2} we prove that if $\alpha_1,\alpha_2,\alpha_3$ are all nonzero, certain transitions are prohibited, and exhibit numerically that all other transitions are allowed. In Section \ref{sec:4.3} we translate the results for the compactification back to the standard symmetric $P_\mathrm{IV}$. In Section \ref{sec:5} we summarize, and present a list of topics for further study. 
	
	\section{Compactification and Fixed Point Analysis}
	\label{sec:compactificationAndEPA}
	In this section we apply a change of variables known as the Poincar\'e compactification to the system \eqref{eq:fsys}. The resulting system is shown to have fourteen fixed points, and the linearization of the system is given at each of the fixed points. 
	
	\subsection{Poincar\'e Compactification}
	 	The (three-dimensional) Poincar\'e compactification consists of two changes of variables: a change of the dependent variables, followed by a change of the independent variable. The former is given by
	 	\begin{subequations}
	 		\label{eq:poincare}
	 		\begin{equation}
	 			\mathbf{g}=\frac{\mathbf{f}}{\sqrt{1+\|\mathbf{f}\|^2}},
	 		\end{equation}
	 		which maps three-dimensional Euclidean space $\mathbf{R}^3$ onto the open unit ball
                        $\mathbf{B}^3:=\left\{ \mathbf{g} \in \mathbf{R}^3:\|\mathbf{g}\|<1 \right\}$. The inverse map is given by
	 		\begin{equation}
	 		\mathbf{f}=\frac{\mathbf{g}}{\sqrt{1-\|\mathbf{g}\|^2}}.
	 		\end{equation}	  
	 	\end{subequations}
 	 In the case of the system \eqref{eq:fsys}, the change of variables \eqref{eq:poincare} results in the system
 	\begin{subequations}
 		\label{eq:gsysx}
	 	\begin{align}
		 	\frac{\mathrm{d} g_1}{\mathrm{d} x}&=\frac{g_1 (g_2-g_3) \left[1-(g_1-g_2)(g_1-g_3) \right]}{\sqrt{1-\|\mathbf{g}\|^2}}+\sqrt{1-\|\mathbf{g} \|^2} \left[ \alpha_1 -g_1 (\alpha_1 g_1+\alpha_2 g_2+\alpha_3 g_3) \right], \\
		 	\frac{\mathrm{d}g_2}{\mathrm{d} x}&=\frac{g_2(g_3-g_1)\left[1-(g_2-g_3)(g_2-g_1) \right]}{\sqrt{1-\|\mathbf{g}\|^2}}+\sqrt{1-\|\mathbf{g}\|^2} \left[ \alpha_2-g_2(\alpha_1 g_1+\alpha_2 g_2+\alpha_3 g_3) \right], \\
		 	\frac{\mathrm{d}g_3}{\mathrm{d} x}&=\frac{g_3(g_1-g_2)\left[1-(g_3-g_1)(g_3-g_2) \right]}{\sqrt{1-\|\mathbf{g}\|^2}}+\sqrt{1-\|\mathbf{g}\|^2} \left[ \alpha_3-g_3(\alpha_1 g_1+\alpha_2 g_2+\alpha_3 g_3) \right].
	 	\end{align}
 	\end{subequations}
        The right hand side here diverges as we approach the ``sphere at infinity'', $\mathbf{S}^2:=\left\{ \mathbf{g} \in \mathbf{R}^3: \| \mathbf{g} \| =1 \right\}$. However if we reparametrize the orbits with a new parameter $t$, defined by 
 	\begin{equation}
 		\label{eq:dxdt}
 		\frac{\mathrm{d} x}{\mathrm{d} t}=\sqrt{1-\|\mathbf{g}\|^2},
 	\end{equation}
	we obtain 
 	\begin{subequations}
 	\label{eq:gsyst}
	\begin{align}
		\frac{\mathrm{d} g_1}{\mathrm{d} t}&=g_1 (g_2-g_3) \left[1-(g_1-g_2)(g_1-g_3) \right]+\left(1-\|\mathbf{g} \|^2 \right) \left[ \alpha_1 -g_1 (\alpha_1 g_1+\alpha_2 g_2+\alpha_3 g_3) \right], \\
		\frac{\mathrm{d}g_2}{\mathrm{d} t}&=g_2(g_3-g_1)\left[1-(g_2-g_3)(g_2-g_1) \right]+\left(1-\|\mathbf{g}\|^2 \right) \left[ \alpha_2-g_2(\alpha_1 g_1+\alpha_2 g_2+\alpha_3 g_3) \right], \\
		\frac{\mathrm{d}g_3}{\mathrm{d} t}&=g_3(g_1-g_2)\left[1-(g_3-g_1)(g_3-g_2) \right]+\left(1-\|\mathbf{g}\|^2 \right) \left[ \alpha_3-g_3(\alpha_1 g_1+\alpha_2 g_2+\alpha_3 g_3) \right].
	\end{align}
	\end{subequations}
	The system \eqref{eq:gsyst} is well-defined on the closed unit ball 
	$\overline{\mathbf{B}^3}:=\left\{ \mathbf{g} \in \mathbf{R}^3:\|\mathbf{g}\|\le 1 \right\}$.
	Furthermore, it is straightforward to check that the sphere at infinity is an invariant manifold of the flow \eqref{eq:gsyst}.

    \subsection{Fixed Point Analysis}

	The system \eqref{eq:gsyst} has $6+6+2=14$ fixed points, all of which lie on the sphere at infinity $\mathbf{S}^2$. We label and categorize them into three types in Table \ref{tab:equilibria}.  In Table \ref{tab:eigen} we give the eigenvalues and eigenvectors of the linearizations at each of the fixed points. 

	\begin{table}
		\begin{center}
			\begin{tabular}{ |c|c|c| } 
				\hline
				Type $A$ & Type $B$ & Type $C$ \\ \hline 
				$A_1^\pm:=\pm \frac{1}{\sqrt{2}} \left(0,-1,1 \right)$ & $B_1^\pm:=\pm (1,0,0) $ & $C^\pm:=\pm \frac{1}{\sqrt{3}} (1,1,1)$ \\ 
				$A_2^\pm:=\pm \frac{1}{\sqrt{2}} \left(1,0,-1 \right)$ & $B_2^\pm:=\pm (0,1,0) $& \phantom{} \\ 
				$A_3^\pm:=\pm \frac{1}{\sqrt{2}} \left(-1,1,0 \right)$ & $B_3^\pm:=\pm (0,0,1)$ & \phantom{} \\
				\hline
			\end{tabular}
		\end{center}
		\caption{Fixed points of the compactified system \eqref{eq:gsyst}.}
		\label{tab:equilibria}
	\end{table}

	\begin{table}
	\begin{center}
		\bgroup
		\def\arraystretch{1.5}
		\begin{tabular}{ |c|c|c| } 
			\hline
			Pair of fixed points & Eigenvalues & Eigenvectors \\ \hline 
			$A_1^\pm $ & $\mp \left\{\frac{3}{\sqrt{2}},\sqrt{2},\frac{1}{\sqrt{2}} \right\}$ & $\begin{bmatrix} -2 \\ 1 \\ 1 \end{bmatrix},\begin{bmatrix} -4\alpha_1 \\ 1+2\alpha_1 \\ 3+2\alpha_1 \end{bmatrix}, \begin{bmatrix} 0 \\ 1 \\ 1  \end{bmatrix}$ \\ 
			$A_2^\pm$ & $\mp \left\{ \frac{3}{\sqrt{2}},\sqrt{2},\frac{1}{\sqrt{2}} \right\} $& $ \begin{bmatrix} 1 \\ -2 \\ 1 \end{bmatrix},\begin{bmatrix} 3+2\alpha_2 \\ -4\alpha_2 \\ 1+2\alpha_2 \end{bmatrix}, \begin{bmatrix} 1 \\ 0 \\ 1  \end{bmatrix}$ \\ 
			$A_3^\pm$ & $\mp \left\{ \frac{3}{\sqrt{2}},\sqrt{2},\frac{1}{\sqrt{2}} \right\}$ & $\begin{bmatrix} 1 \\ 1 \\ -2 \end{bmatrix},\begin{bmatrix} 1+2\alpha_3 \\ 3+2\alpha_3 \\ -4\alpha_3 \end{bmatrix}, \begin{bmatrix} 1 \\ 1 \\ 0  \end{bmatrix}$ \\
			$B_1^\pm $ & $\mp \left\{1,-1,0 \right\} $ & $\begin{bmatrix} 0 \\ 1 \\ 0 \end{bmatrix},\begin{bmatrix} 0 \\ 0 \\ 1 \end{bmatrix}, \begin{bmatrix} 1 \\ -2\alpha_2 \\ 2\alpha_3  \end{bmatrix}$ \\ 
			$B_2^\pm$ & $\mp \left\{1,-1,0 \right\} $ & $\begin{bmatrix} 0 \\ 0 \\ 1 \end{bmatrix},\begin{bmatrix} 1 \\ 0 \\ 0 \end{bmatrix}, \begin{bmatrix} 2\alpha_1 \\ 1 \\ -2\alpha_3  \end{bmatrix}$ \\ 
			$B_3^\pm$ & $\mp\left\{1,-1,0 \right\} $ & $\begin{bmatrix} 1 \\ 0 \\ 0 \end{bmatrix},\begin{bmatrix} 0 \\ 1 \\ 0 \end{bmatrix}, \begin{bmatrix} -2\alpha_1 \\ 2\alpha_2 \\ 1  \end{bmatrix}$ \\ 
			$C^\pm$ & $\pm \left\{i,-i,0 \right\} $ & $\begin{bmatrix} \e^{\frac{2 \pi i}{3}} \\ \e^{-\frac{2 \pi i}{3}} \\ 1 \end{bmatrix},\begin{bmatrix} \e^{-\frac{2 \pi i}{3}} \\ \e^{\frac{2 \pi i}{3}} \\ 1 \end{bmatrix}, \begin{bmatrix} 2(\alpha_2-\alpha_3)-1 \\ 2(\alpha_3-\alpha_1)-1 \\ 2(\alpha_1-\alpha_2)-1  \end{bmatrix}$ \\ 
			\hline
		\end{tabular}
		\egroup
	\end{center}
	\caption{Eigenvalues and eigenvectors of the linearization of the system \eqref{eq:gsyst} at its fixed points.}
	\label{tab:eigen}
        \end{table}

            Note that at each of the fixed points, two (out of three) eigenvectors are tangential to the sphere $\mathbf{S}^2$, reflecting the fact that it is an invariant manifold.

            The type $A$ fixed points are hyperbolic: all the eigenvalues at the points $A_1^+,A_2^+,A_3^+$ are negative, and thus they are attractors. All the eigenvalues at  the points $A_1^-,A_2^-,A_3^-$ are positive and thus they are repellers.

            The type $B$ points all have one positive, one negative and one zero eigenvalue, with the positive and negative eigenvalues associated with eigenvectors that are tangential to the sphere  $\mathbf{S}^2$. From the relation \eqref{eq:fsum} we know that the motion tends towards the sphere if $g_1+g_2+g_3>0$ and away from the sphere if $g_1+g_2+g_3<0$. Thus the points $B_1^+,B_2^+,B_3^+$ ($B_1^-,B_2^-,B_3^-$) each have a two-dimensional stable (unstable) manifold transverse to the sphere and a one-dimensional unstable (stable) manifold on the sphere. The two-dimensional manifolds are made up of one-parameter families of non-generic solutions of the system. 

            The type $C$ points are totally non-hyperbolic, with all eigenvalues having zero real part. The eigenvalues associated with the motion on the sphere are $\pm i$, so in the context of the motion on the sphere at infinity the type $C$ points are either centers or weak foci. In the next section we will study the motion restricted to the sphere, and see that in this context the type $C$ points  are centers. But then in Section \ref{sec:4.1} we will see that despite this, all orbits close to the $C^+$ ($C^-$) point but inside the sphere are attracted (repelled) to (from) the point, so in this sense the $C^+$ ($C^-$) point is an attractor (repeller). 
	
	\section{Dynamics on the Sphere at Infinity}
	\label{sec:infsphere}
	
		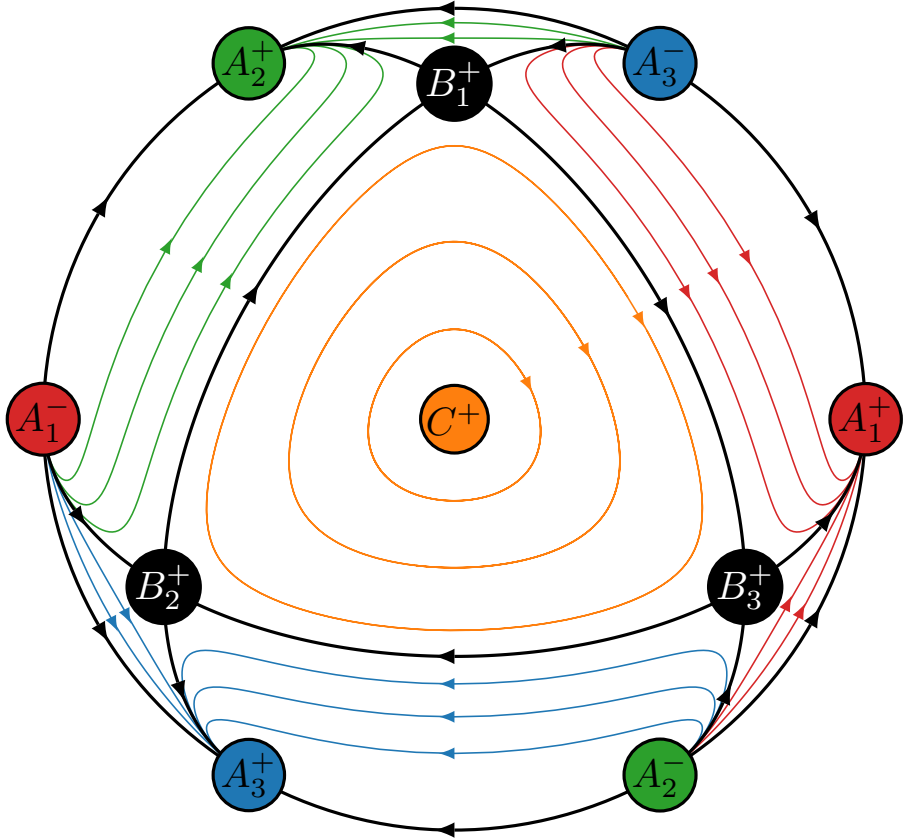
\begin{figure}
			\centering
			\tikzset{mark size=0.5}
				\begin{tikzpicture}[every node/.style={thick,circle,inner sep=0pt,minimum size=0.6cm,fill=white},scale=1.5]
					\tikzmath{\n = 6; \radius =4; \fac = 0.816497; \margin =5; } 
		\begin{axis}[xmin=-4.5,xmax=4.5,ymin=-4.5,ymax=4.5,scale only axis=true, 	width=0.5\textwidth,height=0.5\textwidth,axis lines=none]

		\addplot[mygreen] table [x=a, y=b, col sep=comma, mark=none] {N1.csv};			
		\addplot[mygreen] table [x=a, y=b, col sep=comma, mark=none] {N2.csv};	
		
		\addplot[myred] table [x=a, y=b, col sep=comma, mark=none] {NE1.csv};
		\addplot[myred] table [x=a, y=b, col sep=comma, mark=none] {NE2.csv};		
		\addplot[myred] table [x=a, y=b, col sep=comma, mark=none] {NE3.csv};
		
		\addplot[myred] table [x=a, y=b, col sep=comma, mark=none] {SE1.csv};			
		\addplot[myred] table [x=a, y=b, col sep=comma, mark=none] {SE2.csv};

		\addplot[myblue] table [x=a, y=b, col sep=comma, mark=none] {SW1.csv};			
		\addplot[myblue] table [x=a, y=b, col sep=comma, mark=none] {SW2.csv};				

		\addplot[myblue] table [x=a, y=b, col sep=comma, mark=none] {S1.csv};
		\addplot[myblue] table [x=a, y=b, col sep=comma, mark=none] {S2.csv};		
		\addplot[myblue] table [x=a, y=b, col sep=comma, mark=none] {S3.csv};			
		
		\addplot[mygreen] table [x=a, y=b, col sep=comma, mark=none] {NW1.csv};
		\addplot[mygreen] table [x=a, y=b, col sep=comma, mark=none] {NW2.csv};		
		\addplot[mygreen] table [x=a, y=b, col sep=comma, mark=none] {NW3.csv};		
		
		\addplot[myorange] table [x=a, y=b, col sep=comma, mark=none] {C1.csv};
		\addplot[myorange] table [x=a, y=b, col sep=comma, mark=none] {C2.csv};		
		\addplot[myorange] table [x=a, y=b, col sep=comma, mark=none] {C3.csv};
						
		
		\draw[>-, >=latex,thick] ({360/\n * 0.5}:\radius) 
			arc ({360/\n * 0.5}:{360/\n * 0}:\radius);
		\draw[-, >=latex,thick] ({360/\n * 0.5}:\radius) 
			arc ({360/\n * 0.5}:{360/\n * (1)}:\radius);	
			
		\draw[-, >=latex,thick] ({360/\n * 1.5}:\radius) 
			arc ({360/\n * 1.5}:{360/\n * 1}:\radius);	
		\draw[>-, >=latex,thick] ({360/\n * 1.5}:\radius) 
			arc ({360/\n * 1.5}:{360/\n * (2)}:\radius);
			
		\draw[>-, >=latex,thick] ({360/\n * 2.5}:\radius) 
			arc ({360/\n * 2.5}:{360/\n * 2}:\radius);
		\draw[-, >=latex,thick] ({360/\n * 2.5}:\radius) 
			arc ({360/\n * 2.5}:{360/\n * 3}:\radius);	
			
		\draw[-, >=latex,thick] ({360/\n * 3.5}:\radius) 
			arc ({360/\n * 3.5}:{360/\n * 3}:\radius);
		\draw[>-, >=latex,thick] ({360/\n * 3.5}:\radius) 
			arc ({360/\n * 3.5}:{360/\n * 4}:\radius);	
				
		\draw[>-, >=latex,thick] ({360/\n * 4.5}:\radius) 
			arc ({360/\n * 4.5}:{360/\n * 4}:\radius);
		\draw[-, >=latex,thick] ({360/\n * 4.5}:\radius) 
			arc ({360/\n * 4.5}:{360/\n * 5}:\radius);		
			
		\draw[-, >=latex,thick] ({360/\n * 5.5}:\radius) 
			arc ({360/\n * 5.5}:{360/\n *5}:\radius);
		\draw[>-, >=latex,thick] ({360/\n * 5.5}:\radius) 
			arc ({360/\n * 5.5}:{360/\n * 6}:\radius);	
			
		\draw[>-,>=latex,mygreen] (0,3.71) -- (-0.01,3.71);			
		\draw[>-,>=latex,mygreen] (0,3.86) -- (-0.01,3.86);
		
		\draw[>-,>=latex,myred] (2.81775, 1.62683) -- (2.81775+.01, 1.62683-.02);			
		\draw[>-,>=latex,myred] (2.509, 1.44857) -- (2.509+.01, 1.44857-.02);				
		\draw[>-,>=latex,myred] (2.23164, 1.28844) -- (2.23164+.01, 1.28844-.02);						

		\draw[>-,>=latex,myred] ( 3.21295, -1.855) -- ( 3.21295+.01, -1.855+.02);	\draw[>-,>=latex,myred] (3.34286, -1.93) -- (3.34286+.01, -1.93+.02);	

		\draw[>-,>=latex,myblue] ( -3.21295, -1.855) -- ( -3.21295+.01, -1.855-.02);	\draw[>-,>=latex,myblue] (-3.34286, -1.93) -- (-3.34286+.01, -1.93-.02);				
		
		\draw[>-,>=latex,myblue] (0, -3.25366) -- (0-.01, -3.25366);			
		\draw[>-,>=latex,myblue] (0, -2.89715) -- (0-.01, -2.89715);				
		\draw[>-,>=latex,myblue] (0, -2.57687) -- (0-.01, -2.57687);	

		\draw[>-,>=latex,mygreen] (-2.81775, 1.62683) -- (-2.81775+.01, 1.62683+.02);			
		\draw[>-,>=latex,mygreen] (-2.509, 1.44857) -- (-2.509+.01, 1.44857+.02);				
		\draw[>-,>=latex,mygreen] (-2.23164, 1.28844) -- (-2.23164+.01, 1.28844+.02);				
		
		\draw[>-,>=latex,myorange] (0.688211, 0.397339) -- (0.688211+.01, 0.397339-.02);
		\draw[>-,>=latex,myorange] (1.25267, 0.723231) -- (1.25267+.01, 0.723231-.02);
		\draw[>-,>=latex,myorange] (1.78102, 1.02827) -- (1.78102+.01, 1.02827-.02);
		
		\addplot [domain=0:22.5,samples=50,thick,-,>=latex] ({\radius*0.5*(cos(x)+sin(x))},{\radius*(-3*cos(x)+sin(x))/(2*sqrt(3))});
		\addplot [domain=22.5:45,samples=50,thick,>-,>=latex] ({\radius*0.5*(cos(x)+sin(x))},{\radius*(-3*cos(x)+sin(x))/(2*sqrt(3))});
		\addplot [domain=45:90,samples=50,thick,-<,>=latex] ({\radius*0.5*(cos(x)+sin(x))},{\radius*(-3*cos(x)+sin(x))/(2*sqrt(3))});
		\addplot [domain=90:135,samples=50,thick,-,>=latex] ({\radius*0.5*(cos(x)+sin(x))},{\radius*(-3*cos(x)+sin(x))/(2*sqrt(3))});	
		\addplot [domain=135:153,samples=50,thick,-,>=latex] ({\radius*0.5*(cos(x)+sin(x))},{\radius*(-3*cos(x)+sin(x))/(2*sqrt(3))});			
		\addplot [domain=153:180,samples=50,thick,>-,>=latex] ({\radius*0.5*(cos(x)+sin(x))},{\radius*(-3*cos(x)+sin(x))/(2*sqrt(3))});
		
		\addplot [domain=0:22.5,samples=50,thick,-,>=latex] ({\radius*0.5*(cos(x)-sin(x))},{\radius*(3*cos(x)+sin(x))/(2*sqrt(3))});
		\addplot [domain=22.5:45,samples=50,thick,>-,>=latex] ({\radius*0.5*(cos(x)-sin(x))},{\radius*(3*cos(x)+sin(x))/(2*sqrt(3))});
		\addplot [domain=45:90,samples=50,thick,-<,>=latex] ({\radius*0.5*(cos(x)-sin(x))},{\radius*(3*cos(x)+sin(x))/(2*sqrt(3))});
		\addplot [domain=90:135,samples=50,thick,-,>=latex] ({\radius*0.5*(cos(x)-sin(x))},{\radius*(3*cos(x)+sin(x))/(2*sqrt(3))});
		\addplot [domain=135:153,samples=50,thick,-,>=latex] ({\radius*0.5*(cos(x)-sin(x))},{\radius*(3*cos(x)+sin(x))/(2*sqrt(3))}); 
		\addplot [domain=153:180,samples=50,thick,>-,>=latex] ({\radius*0.5*(cos(x)-sin(x))},{\radius*(3*cos(x)+sin(x))/(2*sqrt(3))}); 
		
		\addplot [domain=0:22.5,samples=50,thick,-,>=latex] ({\radius*(-cos(x))},{\radius*(-sin(x))/(sqrt(3))});
		\addplot [domain=22.5:45,samples=50,thick,>-,>=latex] ({\radius*(-cos(x))},{\radius*(-sin(x))/(sqrt(3))});
		\addplot [domain=45:90,samples=50,thick,-<,>=latex] ({\radius*(-cos(x))},{\radius*(-sin(x))/(sqrt(3))});
		\addplot [domain=90:135,samples=50,thick,-,>=latex] ({\radius*(-cos(x))},{\radius*(-sin(x))/(sqrt(3))});
		\addplot [domain=135:153,samples=50,thick,-,>=latex] ({\radius*(-cos(x))},{\radius*(-sin(x))/(sqrt(3))});
		\addplot [domain=153:180,samples=50,thick,>-,>=latex] ({\radius*(-cos(x))},{\radius*(-sin(x))/(sqrt(3))});
		
		\node[draw, circle,thick,fill=myred] at ({360/\n *0}:\radius) {$A_1^+$};
		\node[draw, circle,thick,fill=myblue] at ({360/\n * 1}:\radius) {$A_3^-$};
		\node[draw, circle,thick,fill=mygreen] at ({360/\n * 2}:\radius) {$A_2^+$};
		\node[draw, circle,thick,fill=myred] at ({360/\n * 3}:\radius) {$A_1^-$};
		\node[draw, circle,thick,fill=myblue] at ({360/\n * (4)}:\radius) {$A_3^+$};
		\node[draw, circle,thick,fill=mygreen] at ({360/\n * 5}:\radius) {$A_2^-$};
		\node[draw, circle,thick,fill=black,text=white] at ({-30}:\fac*\radius) {$B_3^+$};
		\node[draw, circle,thick,fill=black,text=white] at ({90}:\fac*\radius) {$B_1^+$};
		\node[draw, circle,thick,fill=black,text=white] at ({210}:\fac*\radius) {$B_2^+$};
		\node[draw, circle,thick,fill=myorange] at (0:0) {$C^+$};
		
		\end{axis}
		\end{tikzpicture}
		\caption{Orbits of the compactified system \eqref{eq:gsyst} on the upper hemisphere of the sphere at infinity.}
		\label{fig:uproj}
	\end{figure}

	In this section we study the system \eqref{eq:gsyst} restricted to the sphere at infinity $\mathbf{S}^2$. We will show that the restricted system is exactly solvable by exhibiting conserved quantities in each of the hemispheres $\{g_1+g_2+g_3>0\}$  and $\{g_1+g_2+g_3<0\}$ 
	
	Using the  orthogonal change of variables given by
	\begin{align*}
	h_1&=\frac{-g_2+g_3}{\sqrt{2}}, \\
	h_2&=\frac{2g_1-g_2-g_3}{\sqrt{6}}, \\
	h_3&=\frac{g_1+g_2+g_3}{\sqrt{3}},
	\end{align*}
	and defining  spherical coordinates by 
	\begin{align*}
	h_1&=\cos \phi \sin \theta, \\
	h_2&=\sin \phi \sin \theta, \\
	h_3&= \cos \theta,
	\end{align*}
	the flow on the sphere at infinity takes the simple form 	
	\begin{subequations}
		\label{eq:dynsph}
		\begin{align}
			\dot{\phi}&=-\frac{\sin \theta \sin 3 \phi }{\sqrt{2}}-\cos \theta ,  \\
			\dot{\theta}&=\frac{\sin ^2\theta  \cos \theta  \cos 3 \phi }{\sqrt{2}}.
		\end{align}
	\end{subequations} 
	The system \eqref{eq:dynsph} has the conserved quantity
	\begin{equation*}
	 H = 3 \tan^2 \theta+\sqrt{2} \sin 3\phi \tan^3 \theta.
	\end{equation*}
	Note however that this is not defined on the equator of the sphere, $\{\theta=\frac\pi2\} = \{h_3=0\}$. So in fact we have two conserved quantities, one on the open upper hemisphere, the other on the open lower hemisphere. The level sets of $H$ on the upper hemisphere (orbits of the system \eqref{eq:dynsph})  are displayed in Figure \ref{fig:uproj}. The type $C$ point is a center, the type $B$ points are saddles, and the type $A$ points are nodes. Nodes are not allowed in a planar system with a conserved quantity. But $H$ is not defined on the equator, where the type $A$ points are located, so this is not a problem. 

        \section{Generic Orbits of $P_\mathrm{IV}$} 
	
	Having discussed the motion on the sphere at infinity in Section \ref{sec:infsphere}, in this section we return to the study of generic solutions of symmetric $P_\mathrm{IV}$ \eqref{eq:fsys} and its Poincar\'e compactification, \eqref{eq:gsyst}. Because solutions of \eqref{eq:fsys} satisfy \eqref{eq:fsum}, all orbits of \eqref{eq:gsyst} must ``start'' on the  lower hemisphere at infinity ($g_1^2+g_2^2+g_3^2=1$ with $g_1+g_2+g_3 \le 0$) and ``end'' on the upper hemisphere at infinity ($g_1^2+g_2^2+g_3^2=1$ with $g_1+g_2+g_3 \ge 0$). Clearly, one possibility is that orbits tend (as $t \to \pm \infty$) to fixed points on the sphere at infinity. But we have seen in the last section that there are also closed periodic orbits on the sphere at infinity around the type $C$ fixed points, and it is imaginable that orbits inside the sphere might tend (in an orbital sense) to these closed periodic orbits. In Section \ref{sec:4.1} we eliminate this possibility for generic orbits by showing that the type $C^+$ fixed point attracts orbits starting close to it but inside the sphere (and similarly the type $C^-$ fixed point repels). We deduce that {\em generic orbits of \eqref{eq:gsyst} start at one of the type $A^-$ or type $C^-$ fixed points  and end at one of the type $A^+$ or type $C^+$ fixed points}. (The type $B$ fixed points are of mixed stability, and there are non-generic orbits starting and ending at these points.) In Section \ref{sec:4.2} we ask the question whether transitions can occur between all of the type $A^-$ and type $C^-$ fixed points and all of the type $A^+$ and type $C^+$ fixed points. We deduce that there are transition rules, depending on the signs of the parameters $\alpha_i$. In Section \ref{sec:4.3} we discuss the consequences of these results for compactified, symmetric $P_\mathrm{IV}$ for the standard, symmetric $P_\mathrm{IV}$, and the relationship with some existing work on standard $P_\mathrm{IV}$. 
	
	\subsection{Asymptotic Stability of the Type $C^+$ Fixed Point}
	\label{sec:4.1}
	
	The title of this section is slightly misleading, as we have seen that there are closed periodic orbits around the type $C^+$ fixed point on the sphere at infinity. We mean that orbits starting sufficiently close to the type $C^+$ fixed point and strictly inside the sphere all tend to the fixed point.
	
	We recall from Section \ref{sec:compactificationAndEPA} that the linearization at the type $C^+$ fixed point $(g_1,g_2,g_3) = \left(\frac1{\sqrt3} , \frac1{\sqrt3} , \frac1{\sqrt3} \right)$ has eigenvalues $0,\pm i$. Introducing the following linear combinations
	\begin{eqnarray}
	Z  &=&    g_1 + g_2 + g_3 - \sqrt{3},  \nonumber \\
	X  &=&    2(\alpha_2-\alpha_3) Z + 2g_1 - g_2 -  g_3,   \label{XYZ}  \\
	Y  &=&   \frac2{\sqrt3} (1-3\alpha_1) Z + \frac{ g_2 - g_3}{\sqrt3} ,     \nonumber
	\end{eqnarray}
	the system near the fixed points takes the form
	\begin{eqnarray}
	\dot{X} &=&  Y  + \ldots, \nonumber \\   
	\dot{Y} &=&  -X  + \ldots , \label{eq:XYZsystem} \\
	\dot{Z} &=& \frac43 Z^2  + \left(\alpha_1 - \frac13 \right) ZX + \frac{\alpha_2-\alpha_3}{\sqrt3} ZY + \ldots\ .  \nonumber
	\end{eqnarray}
	Here, in the equations for $\dot{X}$ and $\dot{Y}$, the unwritten terms are terms that are quadratic, cubic and quartic in $X,Y,Z$; 
	in the equation for $\dot{Z}$ the unwritten terms are only cubic and quartic. We recall 
        some basic results about perturbations of the harmonic oscillator from, for example \cite{schlomiuk1993algebraic,pearson}. 
	For the system 
	\begin{eqnarray*}
		\dot{X} &=&  Y  + P_2(X,Y) + P_3(X,Y)+ \ldots, \\
		\dot{Y} &=&  -X  + Q_2(X,Y) + Q_3(X,Y) + \ldots,  
	\end{eqnarray*}
	where $P_2,Q_2$ are quadratic terms in $X,Y$, $P_3,Q_3$ are cubic and so on, it is possible to find, term-by-term,  a formal quantity
	$$
	C = \frac12  (X^2+Y^2) + C_3(X,Y) + C_4(X,Y) + \ldots,
	$$
	with $C_3$ cubic and so on, such that $C$ obeys
	$$
	\dot{C} =  V_1 C^2 + V_2 C^3 + \ldots\ . 
	$$
	Here $V_1,V_2,\ldots$ are constants depending on the coefficients of the system. (In fact this result is a slight variant of one
	that appears in \cite{schlomiuk1993algebraic,pearson} but its proof is identical.) The sign of $V_1$ is critical. If $V_1<0$ then the squared amplitude of the
	oscillation, $C$, decays with $t$, behaving, for large $t$  as $\frac1{t}$. In this case the fixed point is stable. If $V_1>0$ then the amplitude grows with time and the fixed point is unstable. If $V_1=0$ then the sign of $V_2$ becomes important: The fixed point is stable if $V_2<0$, with $C$ decaying as $\frac1{\sqrt{t}}$. Moving to our system \eqref{eq:XYZsystem}, the analogous result is that we can find
	formal quantities
	\begin{eqnarray*}
	C &=& \frac12  (X^2+Y^2) + C_3(X,Y,Z) + C_4(X,Y,Z) + \ldots, \\
	K &=&  Z +  K_2(X,Y,Z) + K_3(X,Y,Z) + \ldots,   
	\end{eqnarray*}
	such that 
	\begin{subequations}
	\label{CKsys}
	\begin{align}
		\dot{C} \; &= \; \frac{2}{3\sqrt3} C^2 + \frac43 KC + q_1  K^2 C  + \ldots,   \\ 
		\dot{K} \; &= \; \frac43 K^2 + \frac{2}{3\sqrt3}KC + \frac{4}{3\sqrt3} K^3 + q_2 K^4 + q_3 K^2 C  + \ldots\  \ . 
	\end{align}
	\end{subequations}
	Here $q_1,q_2,q_3$ are constants (depending on the parameters $\alpha_i$) that are irrelevant for our purposes. On the right hand side of these equations we have written all terms that are of order 4 or less, taking $K$ to be of order $1$ and $C$ of order $2$ (as its leading terms are quadratic in $X,Y$). But in fact for small $C,K$, in both equations the right hand sides are dominated by the first two terms. Thus to determine the leading-order behavior of $K\approx Z$ and  $C\approx \frac12(X^2+Y^2)$ we solve the system 
	\begin{subequations}
		\label{CKtruncsys}
		\begin{align}
			\dot{C} \; &= \; \frac{2}{3\sqrt3} C^2 + \frac43 KC,  \\
			\dot{K} \; &= \; \frac43 K^2 + \frac{2}{3\sqrt3}KC,   
		\end{align}
		giving
	\end{subequations}
	\begin{equation}
		C(t) = \frac{-C(0)t_0}{t-t_0} \ , \qquad
		K(t) = \frac{-K(0)t_0}{t-t_0} \ .  \label{CKtruncsol}
	\end{equation}
	Here the constants $t_0$, $C(0)$, $K(0)$ are related by 
	$$ 3 \sqrt3   = 2 t_0 \left( C(0) + 2 \sqrt3 K(0) \right)\ .  $$
	Note that $C(0)$ is positive and $K(0)$ is negative. 
	The requirement that the initial point is strictly inside the unit sphere 
	can be checked, for initial points sufficiently close to the fixed point, 
	to correspond to the condition  $C(0) + 2 \sqrt3 K(0)<0$. Thus $t_0$ is negative
        and $C(t),K(t)$ decay. 	Furthermore the solution \eqref{CKtruncsol} to the truncated system \eqref{CKtruncsys}
        can be shown to extend perturbatively to a solution of the full system  \eqref{CKsys} as a Laurent series,  in negative powers of $t-t_0$.  

	From the leading order behavior \eqref{CKtruncsol} we deduce that orbits 
	starting inside the sphere and sufficiently close to the type $C^+$ fixed point do indeed tend to the fixed point, making it an attractor (and similarly, the type $C^-$ fixed point is a repeller).
	Note that for orbits close to the fixed point $X,Y,Z$, defined in \eqref{XYZ}, decay, for large $t$,  as $\frac1{\sqrt t} , \frac1{\sqrt t} , \frac1{t}$ respectively. There is a unique orbit that (asymptotically) has no component in the $X$ and $Y$ directions (i.e. no oscillatory component). This orbit is of some interest, but since
        it is a non-generic orbit, it will not be investigated in detail in this paper.
	
	There is another way to demonstrate the asymptotic stability of the type $C^+$ fixed point, though it is more of a demonstration, rather than a systematic  derivation as  given above. 
	We start from symmetric, non-compactified  $P_\mathrm{IV}$, the system \eqref{eq:fsys}. It is well known that this has
	{\em Painlev\'e series} (or pole series), such as
	\begin{subequations}
	\label{eq:a3pseries}
	\begin{align}
	  f_1 &= \frac1{x-x_0} +\frac{x_0}{2}+\frac{\frac{x_0^2}{4}+2\alpha_1+\alpha_2+3\alpha_3}{3}\left(x-x_0 \right)
                 + \frac{C+\frac{x_0}{2}(\alpha_1+\alpha_2+3\alpha_3)}{4}(x-x_0)^2 +\ldots   \\
          f_2 &= -\frac1{x-x_0} +\frac{x_0}{2}+\frac{-\frac{x_0^2}{4}+\alpha_1+2\alpha_2+3\alpha_3}{3}\left(x-x_0 \right)
                  +\frac{C-\frac{x_0}{2}(\alpha_1+\alpha_2+3\alpha_3)}{4}(x-x_0)^2  +\ldots   \\
	  f_3 &= -\alpha_3(x-x_0) -\frac{C}{2}(x-x_0)^2+ \ldots  
	\end{align}
	\end{subequations}
	where $x_0,C$ are constants. 
	When translated into a solution of the compactification,
	this series describes the asymptotic behavior of solutions near
	the type $A_3^+$ fixed point as $x$ tends to $x_0$ from below, and solutions near the type $A_3^-$ fixed points as $x$ tends to $x_0$ from
        above. Remarkably, it seems it is possible to define a different type of series solution for
	symmetric $P_\mathrm{IV}$. These series take the following form, very similar to  Lindstedt-Poincar\'e series \cite{drazin1992nonlinear}
	\begin{subequations}
	\label{eq:cseries}
        \begin{eqnarray}
		f_1 &\sim& \frac{x}{3} + a \cos \omega(x)
		+ \frac{\alpha_2-\alpha_3 + \frac12 a^2  \cos 2\omega(x)}{x} \notag \\
		&& + \frac{S_{31}(x)}{x^2} + \frac{S_{41}(x)}{x^3}    + \ldots,   \\
		f_2 &\sim& \frac{x}{3}  -   \frac{a}2    \cos\omega(x) - \frac{\sqrt3 a}{2}  \sin\omega(x)
		+ \frac{\alpha_3-\alpha_1 - \frac14 a^2 \cos 2\omega(x) + \frac{\sqrt3}{4} a^2  \sin 2\omega(x)}{x} \notag \\ 		    
		&& + \frac{S_{32}(x)}{x^2} + \frac{S_{42}(x)}{x^3}    + \ldots,  \\
		f_3 &\sim& \frac{x}{3}  - \frac{a}2    \cos\omega(x) +  \frac{\sqrt3 a}{2}  \sin\omega(x)
		+ \frac{\alpha_1-\alpha_2 - \frac14 a^2 \cos 2\omega(x) - \frac{\sqrt3}{4} a^2  \sin 2\omega(x)}{x} \notag \\ 		    
		&& - \frac{S_{31}(x)+S_{32}(x)}{x^2} - \frac{S_{41}(x)+S_{42}(x)}{x^3}    + \ldots \, . 		    
	\end{eqnarray}
	\end{subequations}
	Here 
	\newline$\bullet$
	$a$ is an arbitrary constant;
	\newline$\bullet$        
	$S_{31}(x), S_{32}(x)$ are odd, third order trigonometric polynomials
	in $\omega(x)$,  i.e. linear combinations of
	\newline $\cos\omega(x) ,\sin\omega(x) ,\cos3\omega(x) ,\sin3\omega(x)$,
	with coefficients determined by $a,\alpha_1,\alpha_2,\alpha_3$; 
	\newline$\bullet$
	$S_{41}(x), S_{42}(x)$ are even, fourth order trigonometric polynomials
	in $\omega(x)$,  i.e. linear combinations of
	\newline $\cos2\omega(x) ,\sin2\omega(x) ,\cos4\omega(x) ,\sin4\omega(x)$ and a constant term,
	with coefficients determined by $a,\alpha_1,\alpha_2,\alpha_3$; 
	\newline$\bullet$   the function $\omega(x)$  satisfies 
	$$ \omega'(x) = \frac{x}{\sqrt3}  - \frac{\sqrt3 a^2}{2x}
	- \frac{\sqrt3\left(3a^4 + 12(\alpha_1^2+\alpha_2^2+\alpha_3^2+\alpha_1) - 8\right) }{4x^3} + \ldots \ .  $$
	\newline
	The coefficients in the series \eqref{eq:cseries}
        can be determined, order-by-order, using a symbolic manipulator.
	There are two free constants, $a$ and $\omega(0)$, in the series, like in the pole series \eqref{eq:a3pseries}. 
	The existence of solutions of $P_\mathrm{IV}$ with asymptotics of this type has been recognized in the literature
        \cite{abdullayev,bassom1992integral,bassom1993numerical,reeger2013painleve}. 
	However the full form of the series given above seems to be new.  
	
	When translated into solutions of compactified symmetric $P_\mathrm{IV}$, these become solutions tending (as $x$ tends to $\infty$) to the type $C^+$ fixed point. For large $x$, the coordinate $t$ grows as $x^2$. Although in the  solutions of the non-compactified system there is a finite amplitude oscillation in the solution for large $x$, in the corresponding solutions of the compactified system the oscillation
	decays as $\frac1{\sqrt{t}}$, as observed. 
	
	Having shown that the $C^+$ point is asymptotically stable, in the sense
	that orbits close to it but strictly inside the sphere are attracted to it,
	we can deduce that generic orbits of \eqref{eq:gsyst} start at one of the type $A^-$
	or type $C^-$ fixed points and end at one of the type $A^+$ or type $C^+$
	fixed points. It remains the case that there may be non-generic orbits that
	tend (in an orbital sense) to specific closed orbits on the sphere at
	infinity, but analyticity of the vector field precludes this for a
	generic set of orbits, and we conjecture this does not happen at all.

	\subsection{Permitted and Forbidden Transmissions}
	\label{sec:4.2}
	It remains to be determined whether there are orbits that connect all of the
	four repeller points to all of the four attractor points. In this section we determine
	the transition rules. We use asymptotic series of the solutions. In \eqref{eq:a3pseries}
	we gave the asymptotic series associated with the type $A_3$ points.
	Here we display the leading terms of the series associated with all the type $A$ points: 
	$$
	\begin{array}{ccc}
	A_1:  &      &  \\
	f_1 & \sim & -\alpha_1(x-x_0) + \ldots \\
	f_2 & \sim & \frac1{x-x_0} + \ldots    \\
	f_3 & \sim & -\frac1{x-x_0} + \ldots 
	\end{array}
	~~~~~~~~
	\begin{array}{ccc}
	A_2: &    &   \\
	f_1 & \sim & -\frac1{x-x_0}   + \ldots \\
	f_2 & \sim & -\alpha_2(x-x_0) + \ldots \\
	f_3 & \sim & \frac1{x-x_0}    + \ldots 
	\end{array}
	~~~~~~~~
	\begin{array}{ccc}  
	A_3: &      &                           \\ 
	f_1 & \sim & \frac1{x-x_0}   + \ldots \\  
	f_2 & \sim & -\frac1{x-x_0}    + \ldots \\
	f_3 & \sim & -\alpha_3(x-x_0) + \ldots   
	\end{array}
	$$
	Note also from \eqref{eq:fsys} that at a regular zero of $f_1$, $f_1'=\alpha_1$, at a regular zero of $f_2$, $f_2'=\alpha_2$, and at a regular zero
	of $f_3$, $f_3'=\alpha_3$.  Here a prime denotes differentiation with respect to $x$, and by a ``regular'' zero, we mean a zero
	at which all $3$ of the functions $f_1,f_2,f_3$ are finite (as opposed to zeros associated with type $A$ and type $B$ fixed points). 
	
	Suppose now that $\alpha_1>0$. Since at a regular zero of $f_1$ we have $f_1'=\alpha_1>0$, it follows that $f_1$ can change from 
	negative to positive without needing to approach a fixed point. From the pole series (equations  above) and the type $C$ series \eqref{eq:cseries}
	we see that
	\begin{center}
		\begin{tabular}{ll}  
			Leaving the type $A_1^-$ point ($x\searrow x_0$), $f_1<0$, &   approaching the type $A_1^+$ point ($x\nearrow x_0$), $f_1>0$. \\
			Leaving the type $A_2^-$ point ($x\searrow x_0$), $f_1<0$, &   approaching the type $A_2^+$ point ($x\nearrow x_0$), $f_1>0$. \\
			Leaving the type $A_3^-$ point ($x\searrow x_0$), $f_1>0$, &   approaching the type $A_3^+$ point ($x\nearrow x_0$), $f_1<0$. \\
			Leaving the type $C^-$    point ($x\rightarrow-\infty$), $f_1<0$, &
			approaching the type $C^+$ point ($x\rightarrow+\infty$), $f_1>0$.  
		\end{tabular}
	\end{center}
	Since between fixed points $f_1$ can only change from negative to positive, we deduce that {\em if $\alpha_1>0$, there cannot be a solution
		connecting the $A_3^-$ and $A_3^+$ points}. 
	
	Now assume $\alpha_1<0$. Then $f_1$ can only change from positive to negative without approaching a fixed point, and we have: 
	\begin{center}
		\begin{tabular}{ll}
			Leaving the type $A_1^-$ point ($x\searrow x_0$), $f_1>0$, &   approaching the type $A_1^+$ point ($x\nearrow x_0$), $f_1<0$. \\
			Leaving the type $A_2^-$ point ($x\searrow x_0$), $f_1<0$, &   approaching the type $A_2^+$ point ($x\nearrow x_0$), $f_1>0$. \\
			Leaving the type $A_3^-$ point ($x\searrow x_0$), $f_1>0$, &   approaching the type $A_3^+$ point ($x\nearrow x_0$), $f_1<0$. \\
			Leaving the type $C^-$     point ($x\rightarrow-\infty$), $f_1<0$, &
			approaching the type $C^+$ point ($x\rightarrow+\infty$), $f_1>0$.  
		\end{tabular}
	\end{center}
	Thus {\em if $\alpha_1<0$ there can be no transitions from either of the $A_2^-$ or $C^-$ points to either of the 
		$A_2^+$ or $C^+$ points}.
	
\begin{figure}
		\centering
		\begin{subfigure}[b]{0.3\textwidth}
			\includegraphics[width=\textwidth]{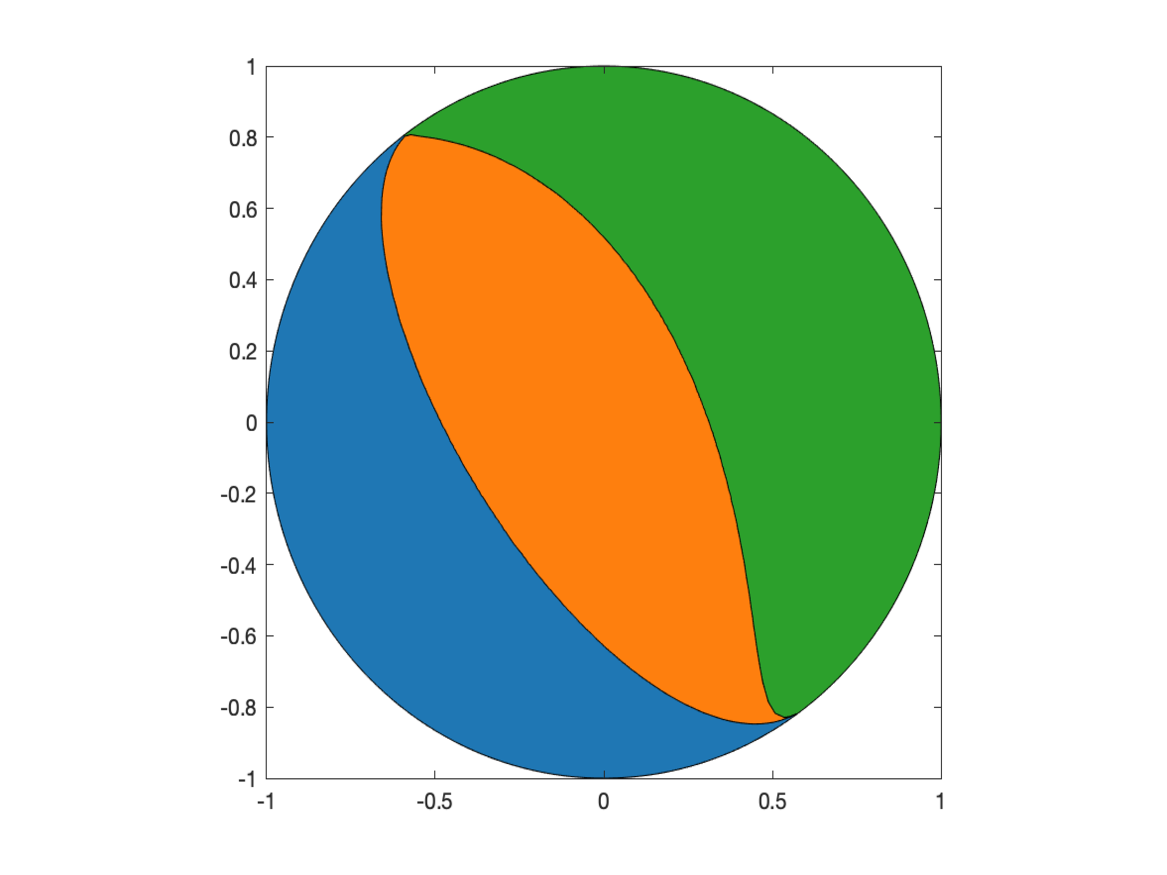}
			\caption{$A_1^-$}
			\label{fig:pppa1}
		\end{subfigure}
		~
		\begin{subfigure}[b]{0.3\textwidth}
			\includegraphics[width=\textwidth]{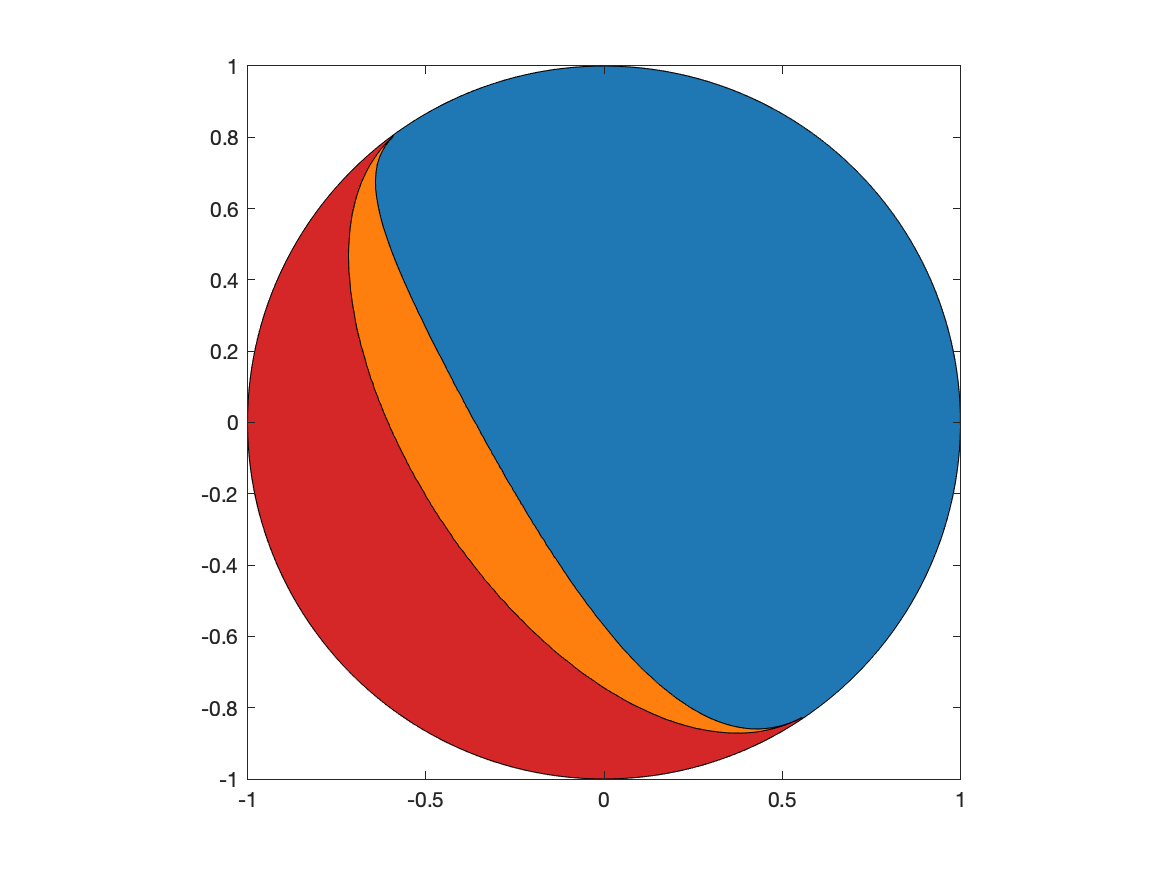}
			\caption{$A_2^-$}
			\label{fig:pppa2}
		\end{subfigure}
		~ 
		\begin{subfigure}[b]{0.3\textwidth}
			\includegraphics[width=\textwidth]{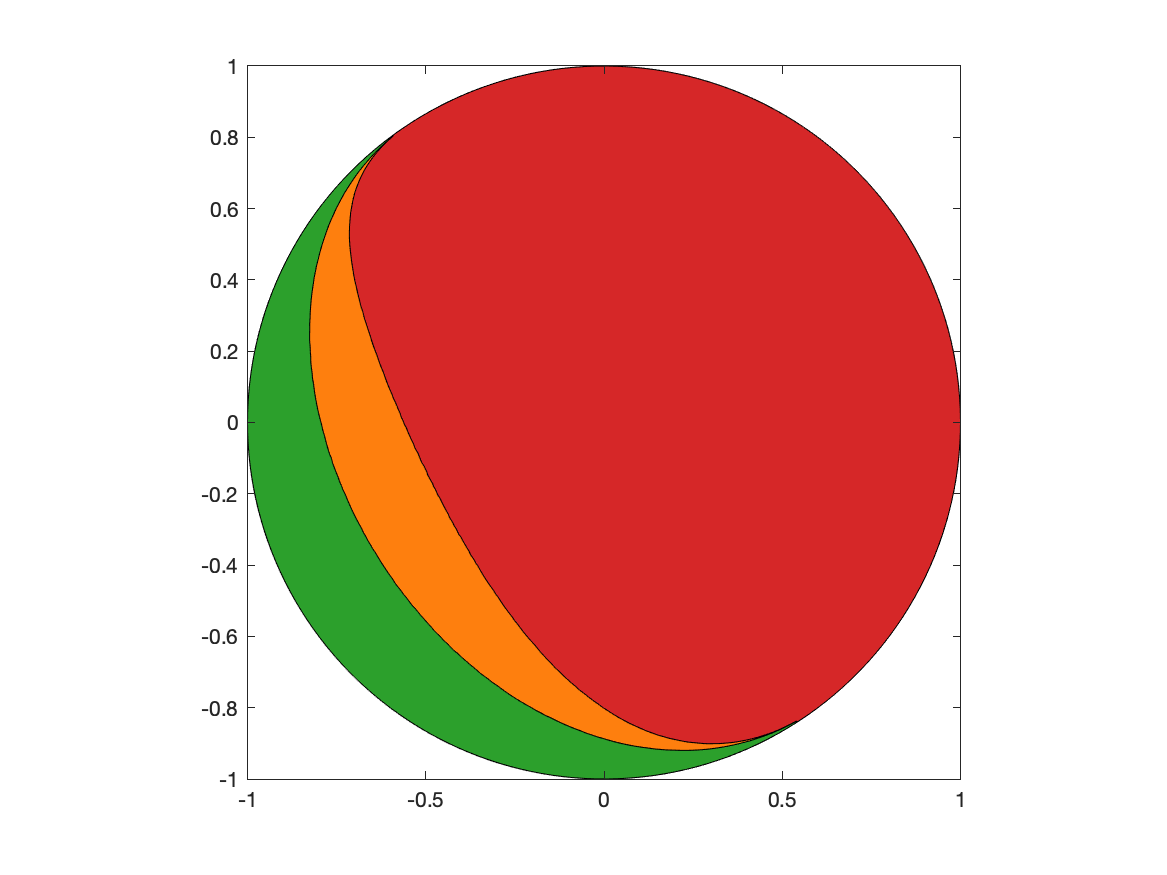}
			\caption{$A_3^-$}
			\label{fig:pppa3}
		\end{subfigure}
		\caption{The fate of orbits starting close to the type $A$ repeller points. In each case
	          we look at orbits starting at points on a small hemispherical cap around the fixed point, and color
                  points according to their destination: $A_1^+$ (red), $A_2^+$ (green), $A_3^+$
	          (blue) or $C$ (orange). Case 1: $\alpha_1=0.2,\alpha_2=0.3,\alpha_3=0.5$.}\label{fig:ppp}
	\end{figure}

\begin{figure}
	\centering
	\begin{subfigure}[b]{0.3\textwidth}
		\includegraphics[width=\textwidth]{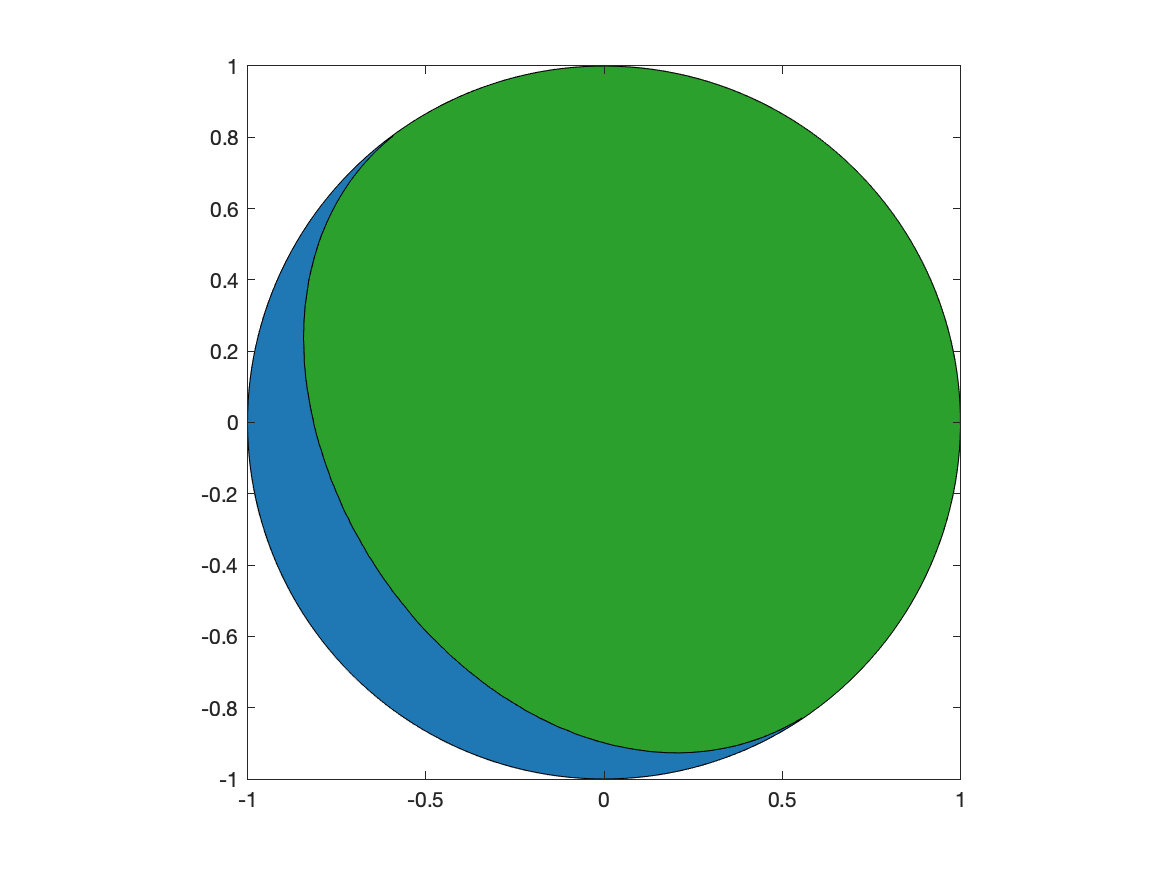}
		\caption{$A_1^-$}
		\label{fig:ppma1}
	\end{subfigure}
	~ 
	\begin{subfigure}[b]{0.3\textwidth}
		\includegraphics[width=\textwidth]{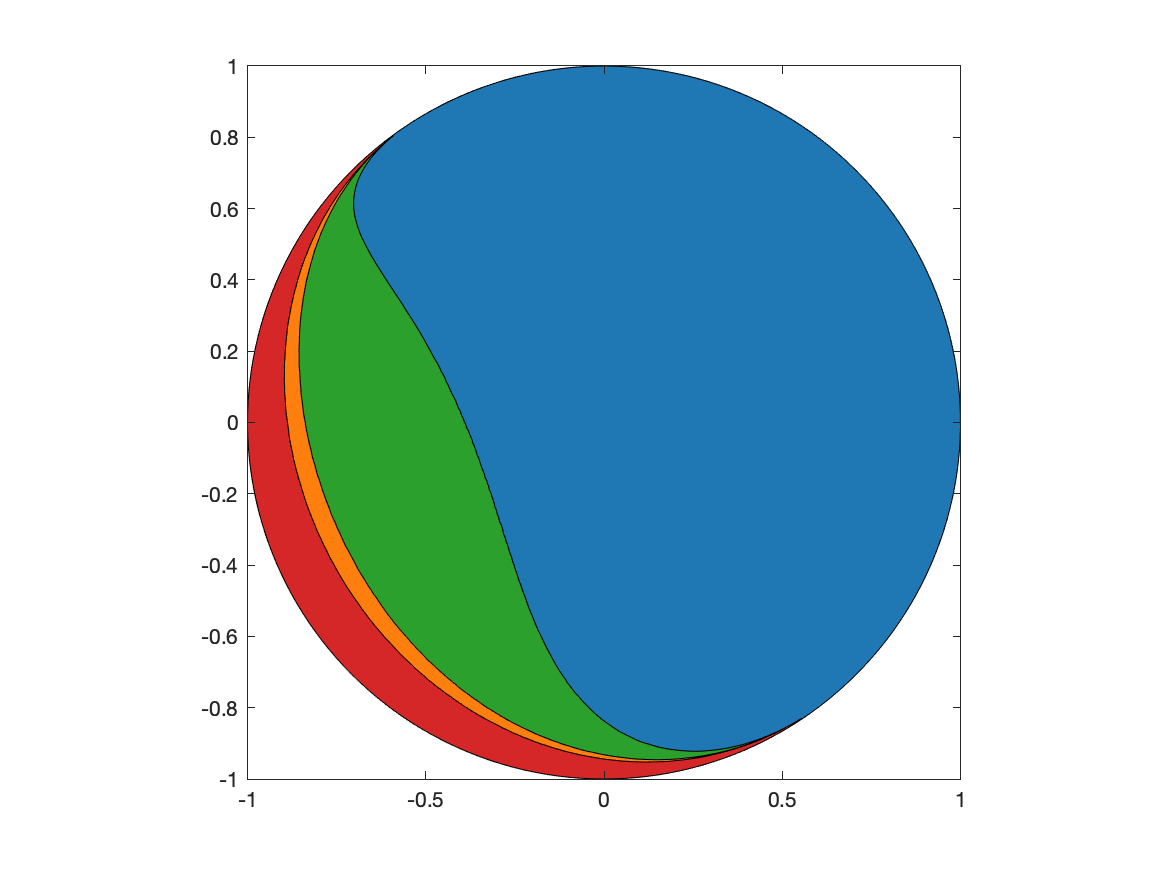}
		\caption{$A_2^-$}
		\label{fig:ppma2}
	\end{subfigure}
	~ 
	\begin{subfigure}[b]{0.3\textwidth}
		\includegraphics[width=\textwidth]{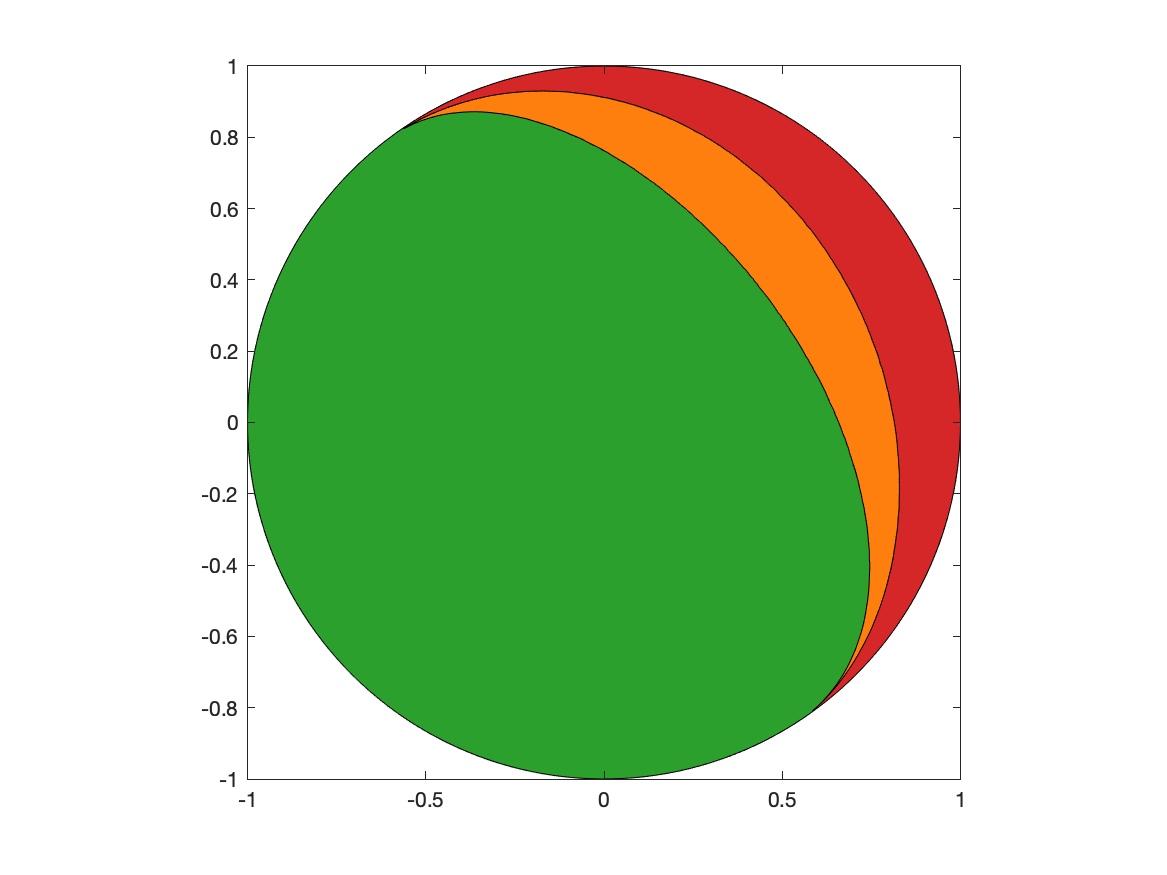}
		\caption{$A_3^-$}
		\label{fig:ppma3}
	\end{subfigure}
	\caption{Case 2: $\alpha_1=0.6,\alpha_2=0.7,\alpha_3=-0.3$.}\label{fig:ppm}
	\end{figure}

\begin{figure}
	\centering
	\begin{subfigure}[b]{0.3\textwidth}
		\includegraphics[width=\textwidth]{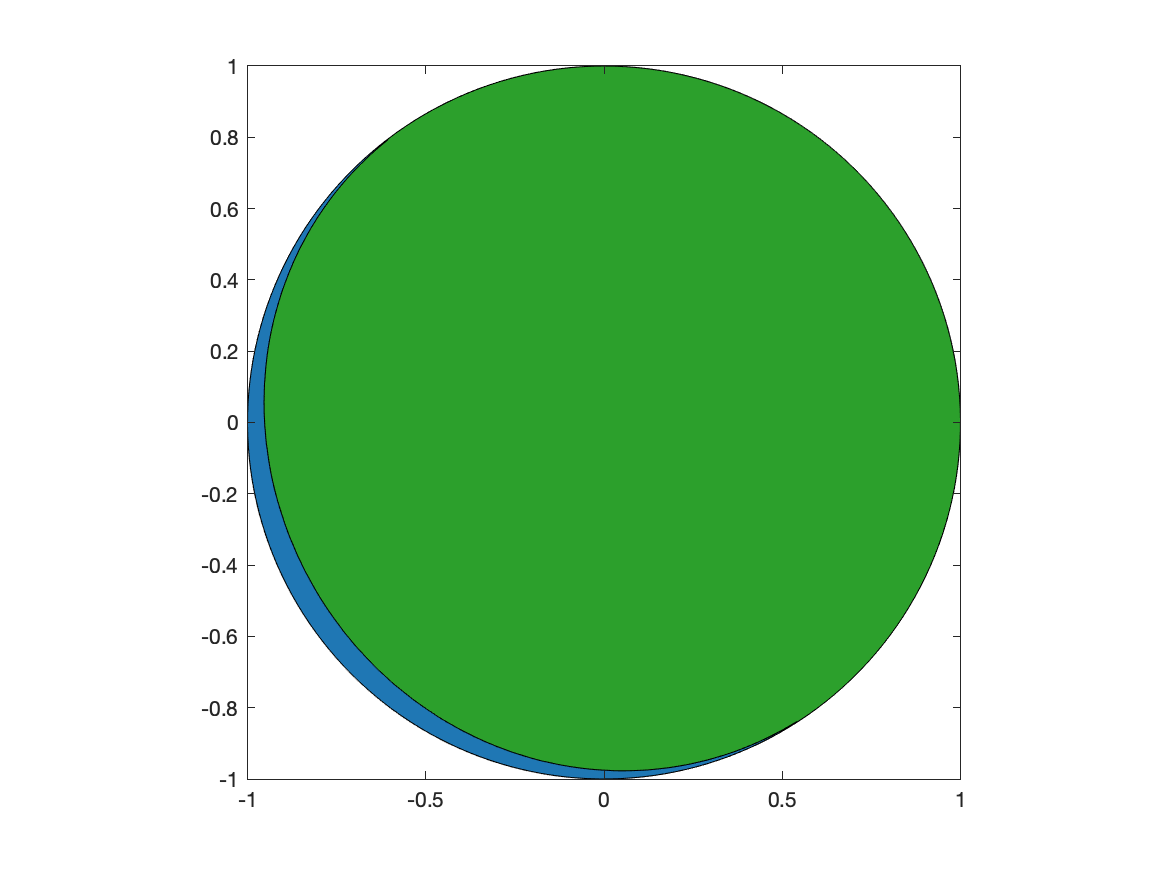}
		\caption{$A_1^-$}
		\label{fig:pmma1}
	\end{subfigure}
	~ 
	\begin{subfigure}[b]{0.3\textwidth}
		\includegraphics[width=\textwidth]{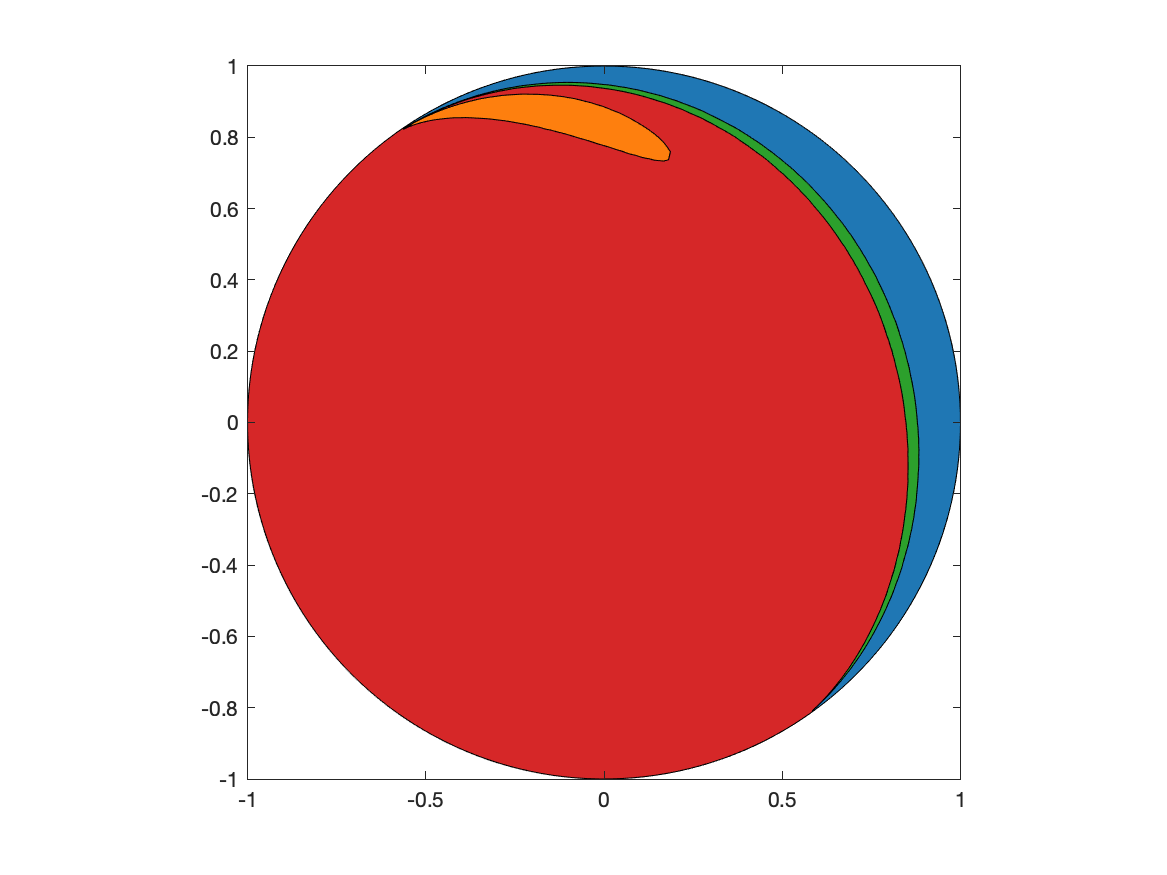}
		\caption{$A_2^-$}
		\label{fig:pmma2}
	\end{subfigure}
	~ 
	\begin{subfigure}[b]{0.3\textwidth}
		\includegraphics[width=\textwidth]{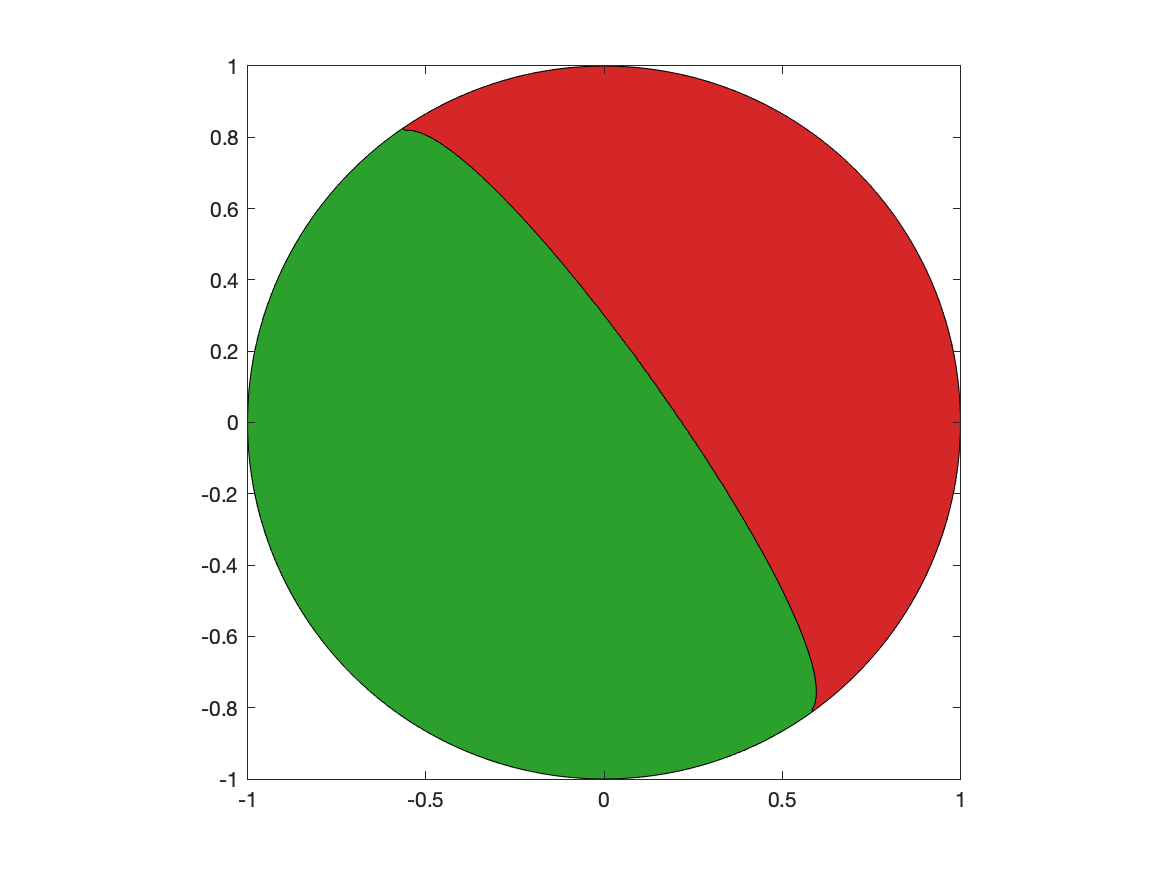}
		\caption{$A_3^-$}
		\label{fig:pmma3}
	\end{subfigure}
	\caption{Case 3: $\alpha_1=1.3,\alpha_2=-0.2,\alpha_3=-0.1$.}\label{fig:pmm}
	\end{figure}

	Similar conclusions can be reached using the signs of $\alpha_2$ and $\alpha_3$.  Note that since $\alpha_1+\alpha_2+\alpha_3=1$,
	there are three possibilities: All $\alpha_i$ positive, one negative and two negative.
	Table \ref{tab:excls} shows the excluded transitions in the various cases. We also verified these results
	numerically. Because of the oscillations, following orbits starting near the $C^-$ point is delicate, so we looked only at
	orbits starting near the type $A^-$ points. For each of the three points, we looked at orbits starting at points on a small hemispherical
	cap around the fixed point, and colored these points according to their destination: $A_1^+$ (red), $A_2^+$ (green), $A_3^+$
	(blue) or $C^+$ (orange). Results for three different choices of the parameters (three positive, two positive, one positive)
        are displayed in Figures \ref{fig:ppp}, \ref{fig:ppm} and \ref{fig:pmm} respectively. It can be checked that the transition rules are all respected. 
	
	\begin{table}
		\begin{center}
			\begin{tabular}{c|cccc}
				$+++$   & $C^+$  & $A_1^+$ & $A_2^+$ &  $A_3^+$  \\ \hline
				$C^-$   &        &            &      &   \\
				$A_1^-$ &        &  \text{\sffamily X} &  & \\
				$A_2^-$ &        &            & \text{\sffamily X}  &  \\ 
				$A_3^-$ &        &            &   &  \text{\sffamily X}
			\end{tabular}
		\end{center}
		\begin{center}
			\begin{tabular}{c|cccc}
				$++-$    & $C^+$  & $A_1^+$ & $A_2^+$ &  $A_3^+$  \\\hline
				$C^-$   & \text{\sffamily X} & \text{\sffamily X}       &       &         \\
				$A_1^-$ & \text{\sffamily X} & \text{\sffamily X}       &       &         \\
				$A_2^-$ &        &       &       &         \\ 
				$A_3^-$ &        &       &       &  \text{\sffamily X}
			\end{tabular}
			~~
			\begin{tabular}{c|cccc}
				$-++$    & $C^+$  & $A_1^+$ & $A_2^+$ &  $A_3^+$  \\ \hline
				$C^-$   & \text{\sffamily X} &       & \text{\sffamily X}     &         \\
				$A_1^-$ &         & \text{\sffamily X}       &       &         \\
				$A_2^-$ & \text{\sffamily X}       &       &  \text{\sffamily X}      &         \\ 
				$A_3^-$ &        &       &       &  
			\end{tabular}
			~~
			\begin{tabular}{c|cccc}
			        $+-+$   & $C^+$  & $A_1^+$ & $A_2^+$ &  $A_3^+$  \\ \hline
				$C^-$   & \text{\sffamily X} &      &       & \text{\sffamily X}       \\
				$A_1^-$ &       &        &       &         \\
				$A_2^-$ &        &       &  \text{\sffamily X}      &         \\ 
				$A_3^-$ & \text{\sffamily X}       &       &       & \text{\sffamily X}
			\end{tabular}
		\end{center}
		\begin{center}
			\begin{tabular}{c|cccc}
				$--+$    & $C^+$     & $A_1^+$ & $A_2^+$  &  $A_3^+$  \\ \hline
				$C^-$   & \text{\sffamily X}  &         &\text{\sffamily X} & \text{\sffamily X} \\
				$A_1^-$ &           &         &         &           \\
				$A_2^-$ & \text{\sffamily X}  &         &\text{\sffamily X} &           \\ 
				$A_3^-$ & \text{\sffamily X}  &         &         &  \text{\sffamily X} 
			\end{tabular}
			~~
			\begin{tabular}{c|cccc}
				$+--$   & $C^+$  & $A_1^+$ & $A_2^+$ &  $A_3^+$  \\ \hline
				$C^-$   & \text{\sffamily X}  & \text{\sffamily X} &           &  \text{\sffamily X}  \\
				$A_1^-$ & \text{\sffamily X}  & \text{\sffamily X} &           &            \\
				$A_2^-$ &           &          &           &            \\ 
				$A_3^-$ & \text{\sffamily X}  &          &           &  \text{\sffamily X} 
			\end{tabular}
			~~
			\begin{tabular}{c|cccc}
				$-+-$   & $C^+$  & $A_1^+$ & $A_2^+$ &  $A_3^+$  \\ \hline
				$C^-$   & \text{\sffamily X}&  \text{\sffamily X}  & \text{\sffamily X} &        \\
				$A_1^-$ & \text{\sffamily X} & \text{\sffamily X}&         &         \\
				$A_2^-$ & \text{\sffamily X} &         & \text{\sffamily X} &         \\ 
				$A_3^-$ &         &         &         &     
			\end{tabular}
		\end{center}
	\caption{
          Excluded transitions:
           Top line: $\alpha_1,\alpha_2,\alpha_3>0$.
           Middle line:  
			$\alpha_1,\alpha_2>0$, $\alpha_3<0$ (left) ; 
			$\alpha_2,\alpha_3>0$, $\alpha_1<0$ (middle) ;
		       $\alpha_3,\alpha_1>0$, $\alpha_2<0$ (right).
           Bottom line: 
			$\alpha_1,\alpha_2<0$, $\alpha_3>0$ (left) ; 
			$\alpha_2,\alpha_3<0$, $\alpha_1>0$ (middle) ;
	   $\alpha_3,\alpha_1<0$, $\alpha_2>0$ (right).  }
                        \label{tab:excls}
        \end{table}
	
	\subsection{Implications for Symmetric $P_\mathrm{IV}$ and standard $P_\mathrm{IV}$}
	\label{sec:4.3}
	
	The results on transition rules presented in the previous section apply to compactified, symmetric $P_\mathrm{IV}$, in which the motion between one
        fixed point and another takes an infinite time $t$. Reverting to non-compactified, symmetric $P_\mathrm{IV}$ or standard
	$P_\mathrm{IV}$,  the motion to (from) a type $A^+$ ($A^-$) point takes only a finite time $x$. However, since such motions end in
        a pole-type singularity, it is possible to continue the motion past the pole.
        A motion ending at the $A_i^+$  fixed point is concatenated with further motion beginning at the $A_i^-$ fixed point. 
	
	So, for example,  in the case that all of the $\alpha_i$ are positive, we see that it is possible to have a type $C$ to type $C$ transition
	of the compactified system, which would give rise to solution of the non-compactified system that is finite for all $x$, with asymptotics
	given by \eqref{eq:cseries} as $x\rightarrow \pm\infty$ (with, in general, different values of the parameters in this series
	for $x\rightarrow+\infty$  and for  $x\rightarrow-\infty$). There is a well-known example of such a solution, the solution $f_1=f_2=f_3$ in the
	case $\alpha_1=\alpha_2=\alpha_3=\frac13$. But in fact we expect a full two-parameter family of such solutions, at least for some set of parameter
	values. We also expect solutions (of non-compactified, symmetric $P_\mathrm{IV}$) 
	with type $C$ behavior as $x\rightarrow\pm\infty$ but with a sequence of pole singularities for finite $x$'s, corresponding to
        passing a finite number of type $A$ points. The transition rules in this case 
	dictate that two singularities of the same type cannot follow each other without a singularity of a different type between them. 
	
	In the case, say,  that  $\alpha_1,\alpha_2$ are negative and $\alpha_3$ is positive, a solution that has type $C$ behavior at
	$\pm\infty$ cannot be finite for all $x$, but must have at least one type $A_1$ singularity. There can be singularities of all
	types, but the transition rules dictate that there cannot be two successive type $A_2$ singularities,
	or two successive type $A_3$ singularities,  and the final singularities, both as $x\rightarrow+\infty$ and as $x\rightarrow-\infty$, 
	must be of type $A_1$. 
	
	We remind that all this discussion pertains only to generic solutions, in particular we have not considered orbits that
	begin or end at type $B$ fixed points. We also remind that the discussion of transition rules has assumed none of the $\alpha_i$
	vanish, which excludes certain cases in which many exact solutions of $P_\mathrm{IV}$ are known. 
	
	We mention connections with some previous work on $P_\mathrm{IV}$. The analytical paper \cite{bassom1992integral} and the numerical paper
	\cite{bassom1993numerical} study certain specific solutions of $P_\mathrm{IV}$, with one of the coefficients $\alpha_i$ vanishing.
	They impose a vanishing boundary condition as $x\rightarrow +\infty$ which corresponds, in the language of this paper,  to looking
	at the one-parameter, non-generic families of solutions that tend to type $B^+$ points. Looking at the asymptotics as $x\rightarrow-\infty$
	they identify a bifurcation; on one side of this bifurcation the solution emanates from the type $C^-$ fixed point (and is finite for
	all $x$), on the other side from one of the type $A^-$ points. The papers \cite{reeger2013painleve,reeger2014painleve}
	use advanced numerical techniques that make it possible to integrate $P_\mathrm{IV}$ through poles, applying this first in the special case
	$\alpha_1=0$, $\alpha_2=\alpha_3=\frac12$, and then for more general parameter values (including cases of $P_\mathrm{IV}$ \eqref{eq:p4} with
	$\beta>0$). Various plots of distributions of poles and zeros on the real axis are shown, some of which are consistent with
	results of this paper (and others are not, as they involve cases in which one or more of the parameters $\alpha_i$ vanish). Solutions
	with asymptotics associated with convergence to the $B^+$ fixed points appear as one-parameter families in the spaces of initial values. 
	In some plots a special solution with apparently unique asymptotic form is indicated; in the language of this paper, this corresponds
	to the unique solution that tends to the $C^+$ fixed point, with zero oscillatory part. 
        
	\section{Summary and Questions for Further Study}
	\label{sec:5}

        We summarize what we have found:  For compactified symmetric $P_\mathrm{IV}$, generic orbits connect one of four repellers to one of four attractors,
        with certain transitions excluded, depending on the signs of the parameters $\alpha_1,\alpha_2,\alpha_3$. For non-compactified, symmetric
        $P_\mathrm{IV}$ this implies that along the real axis generic solutions can have a sequence of poles (and zeros), which maybe be finite,
        infinite in one direction
        or infinite in both directions. If the sequence is finite or infinite in one direction, the asymptotic behavior beyond the singularities is given by
        \eqref{eq:cseries}. Other non-generic solutions display different asymptotic behavior. While none of these behaviors by themselves are new, the dynamical
        systems approach gives a useful perspective on the situation. Also the fact that there are three different types of pole type singularities, and certain
        excluded transitions between them, depending on the signs of the parameters, does not seem to have been fully appreciated. 

        We list a number of questions for further investigation:

        \begin{itemize}
        \item The non-generic orbits need much study, specifically to understand the topology of the stable and unstable manifolds of the type $B^+$ and type
          $B^-$ points respectively, and how this varies with the parameters $\alpha_i$. Here  we just give the asymptotic series for solutions
          of symmetric $P_\mathrm{IV}$ that tend to a type $B_1$ point:
       	\begin{subequations}
	\label{eq:bseries}
        \begin{eqnarray}
	  f_1 &\sim& x + \frac{\alpha_3-\alpha_2}{x} + \frac{a_{102}}{x^3} + \ldots   \notag  \\
         &&  +  \frac{ce^{-\frac12 x^2}}{x^{1+2(\alpha_3-\alpha_2)}}\left(  x + \frac{a_{111}}{x} + \frac{a_{112}}{x^3}  + \ldots \right)      \notag \\
          &&  +  \left( \frac{ce^{-\frac12 x^2}}{x^{1+2(\alpha_3-\alpha_2)}} \right)^2   \left(  - x + \frac{a_{121}}{x} + \frac{a_{122}}{x^3} + \ldots\right)  +  \ldots,   \\
	  f_2 &\sim&  \frac{\alpha_2}{x}  + \frac{a_{202}}{x^3} +\ldots  \notag  \\
         &&  +  \frac{ce^{-\frac12 x^2}}{x^{1+2(\alpha_3-\alpha_2)}}\left(  -x + \frac{a_{211}}{x} + \frac{a_{212}}{x^3}    + \ldots \right)      \notag \\
         &&  +  \left( \frac{ce^{-\frac12  x^2}}{x^{1+2(\alpha_3-\alpha_2)}} \right)^2   \left(   x + \frac{a_{221}}{x} +\frac{a_{222}}{x^3} +\ldots   \right)  + \ldots,   \\
	  f_3 &\sim&   - \frac{\alpha_3}{x}  - \frac{a_{102}+a_{202}}{x^3}   + \ldots \notag  \\
         &&  +  \frac{ce^{-\frac12 x^2}}{x^{1+2(\alpha_3-\alpha_2)}}\left(  -\frac{a_{111}+a_{211}}{x} - \frac{a_{112}+a_{212}}{x^3}  +\ldots  \right)      \notag \\
          &&  +  \left( \frac{ce^{-\frac12 x^2}}{x^{1+2(\alpha_3-\alpha_2)}} \right)^2  
                  \left(  -\frac{a_{121}+a_{221}}{x} - \frac{a_{122}+a_{222}}{x^3}  +\ldots  \right) + \ldots   
	\end{eqnarray}
	\end{subequations}
        Here $c$ is an arbitrary constant, and all the other constants ($a_{102},a_{111},a_{112},a_{121},a_{122},a_{202},a_{211},a_{212},a_{221},a_{222}$ etc.) are determined
        by the the parameters $\alpha_i$. A similar series was written down in \cite{reeger2013painleve}.

        \item  How do the symmetries (or B\"acklund transformations) of $P_\mathrm{IV}$ act upon the picture we have described? Note that certain symmetries
           change the values of the parameters. 
         \item  $P_\mathrm{IV}$, for specific parameter values,  has various families of special solutions, including rational solutions,
           solutions involving the complementary error function, and solutions involving parabolic cylinder functions. All
           of these need to be catalogued according to the sequences of fixed points involved in the dynamical systems picture.  
          \item  Can the results in this paper be extended to  $P_\mathrm{IV}$, equation \eqref{eq:p4},  in the case $\beta>0$? 
          \item  Can the methods of this paper also be applied to other Painlev\'e equations? We note that autonomous dynamical systems that
            are equivalent to other Painlev\'e have appeared in the literature \cite{adler,wh}. Our initial investigations suggest that
            different compactifications may be necessary. 
          \end{itemize}

	\bibliographystyle{alpha}
	\bibliography{p4bib.bib}
		
\end{document}